\documentclass[aps,prl,twocolumn,reprint,groupedaddress,superscriptaddress,floatfix,10pt]{revtex4-1}
\usepackage{ulem}
\usepackage{graphicx}
\usepackage{soul,color}

\begin{document}

\title{Initiating and monitoring the evolution of single electrons within atom-defined structures}

\author{Mohammad Rashidi}
\email[Correspondence to:]{ rashidi@ualberta.net, wyattvine@gmail.com, thdienel@gmail.com}
\affiliation{Department of Physics, University of Alberta, Edmonton, Alberta, T6G 2J1, Canada.}
\affiliation{Nanotechnology Initiative, Edmonton, AB, Canada, T6G 2M9}
\affiliation{Quantum Silicon, Edmonton, AB, Canada, T6G 2M9}

\author{Wyatt Vine}
\email[Correspondence to:]{ rashidi@ualberta.net, wyattvine@gmail.com, thdienel@gmail.com}
\affiliation{Department of Physics, University of Alberta, Edmonton, Alberta, T6G 2J1, Canada.}

\author{Thomas Dienel}
\email[Correspondence to:]{ rashidi@ualberta.net, wyattvine@gmail.com, thdienel@gmail.com}
\affiliation{Department of Physics, University of Alberta, Edmonton, Alberta, T6G 2J1, Canada.}
\affiliation{Nanotechnology Initiative, Edmonton, AB, Canada, T6G 2M9}

\author{Lucian Livadaru}
\affiliation{Quantum Silicon, Edmonton, AB, Canada, T6G 2M9}

\author{Jacob Retallick}
\affiliation{Department of Electrical and Computer Engineering, University of British Columbia, Vancouver, BC, V6T 1Z4, Canada.}

\author{Taleana Huff}
\affiliation{Department of Physics, University of Alberta, Edmonton, Alberta, T6G 2J1, Canada.}
\affiliation{Quantum Silicon, Edmonton, AB, Canada, T6G 2M9}

\author{Konrad Walus}
\affiliation{Department of Electrical and Computer Engineering, University of British Columbia, Vancouver, BC, V6T 1Z4, Canada.}

\author{Robert A. Wolkow}
\affiliation{Department of Physics, University of Alberta, Edmonton, Alberta, T6G 2J1, Canada.}
\affiliation{Nanotechnology Initiative, Edmonton, AB, Canada, T6G 2M9}
\affiliation{Quantum Silicon, Edmonton, AB, Canada, T6G 2M9}

\begin{abstract}
Using a non-contact atomic force microscope we track and manipulate the position of single electrons confined to atomic structures engineered from silicon dangling bonds (DBs) on the hydrogen terminated silicon surface. By varying the probe-sample separation we mechanically manipulate the equilibrium position of individual surface silicon atoms and use this to directly switch the charge state of individual DBs. Because this mechanism is based on short range interactions and can be performed without applied bias voltage, we maintain both site-specific selectivity and single-electron control. We extract the short range forces involved with this mechanism by subtracting the long range forces acquired on a dimer vacancy site. As a result of relaxation of the silicon lattice to accommodate negatively charged DBs we observe charge configurations of DB structures that remain stable for many seconds at 4.5~K. Subsequently we use charge manipulation to directly prepare the ground state and metastable charge configurations of DB structures composed of up to six atoms.

\end{abstract}

\maketitle
Atomic manipulation~\cite{Schweizer1990,Sugimoto2008} has emerged as a powerful strategy to fabricate novel atomic physical-systems~\cite{Slot2016,Drost2017,Folsch2014} and devices~\cite{Khajetoorians2011,Fuechsle2012,Kalff2016,Huff2017a}. An important addition to this experimental toolkit would be the ability to design and control functional atomic charge configurations with single electron precision. To this end, several studies have demonstrated the ability to create, move, and controllably switch single charged species on a surface with scanning probe techniques~\cite{Repp2004,Nazin2005,Gross2009,Sterrer2007,Leoni2011,steurer2015b,Steurer2015probe,Bennett2010,Teichmann2011,Setvin2017,Fatayer2018}. One commonality of prior charge manipulation studies is that they have relied upon the application of bias voltage to induce charge transitions. In most cases this results in a non-negligible tunneling current, whereas in principle charge manipulation could be performed by transferring single electrons. Two recent works highlight progress in this area: Steurer et al. \cite{Steurer2015probe} have demonstrated the lateral manipulation of charge between pentacene molecules adsorbed to a NaCl thin film and Fatayer et al.~\cite{Fatayer2018} have performed charge manipulation with zA tunneling currents.

Building on these efforts we present the manipulation of charge within engineered atomic nanostructures based on single electron events at zero applied bias voltage (0 V). We investigate atom defined charge configurations composed of patterned silicon DBs on a hydrogen-terminated Si(100)-2$\times$1 surface. One advantage to working on the silicon surface is that because DBs are midgap states they are electronically isolated from the bulk substrate~\cite{Haider2009}. They can therefore localize charge without the requirement of a thin insulating film between structure and substrate, which has been essential in many previous studies~\cite{Repp2004,Nazin2005,Gross2009,Sterrer2007,Leoni2011,steurer2015b,Steurer2015probe,Fatayer2018,Liu2015a,Schulz2015a}. The regular spacing of DBs is also guaranteed by the crystal lattice. Recent advances in patterning of DBs now allows large error-free structures to be created~\cite{Pavlicek2017a,Huff2017,Achal2018}. Recent non-contact atomic force microscopy (nc-AFM) measurements~\cite{Rashidi2017} have confirmed that the energy of the negative to neutral charge transition of an isolated DB on a highly n-doped sample is close to the bulk Fermi level (within a few hundred meV). This enables the charge state of DBs to be selectively modified by shifting this charge transition level above or below the bulk Fermi level with bias voltage or other nearby charged DBs~\cite{Huff2017a,Haider2009,Taucer2014,Rashidi2017}. In contrast, here we demonstrate that the charge manipulation of DBs can also be achieved mechanically by using the probe to directly manipulate the equilibrium position of the host atom, making it favorable to host a negative charge. Because this ability is based on short range interactions between the probe and the target atom and can be performed at 0~V we maintain both site-specific selectively and single-electron control.

All experiments were performed on an Omicron LT STM/AFM operating at 4.5~K and ultrahigh vacuum ($<1\times10^{-10}$~Torr). Tips were created from polycrystalline tungsten wire that was chemically etched, sharpened with a focussed ion beam, and attached to a qPlus sensor~\cite{Giessibl2000}. The tips had resonance frequencies of 28~kHz, Q-factors between 12k and 14k, and were driven with an amplitude of 50~pm. Frequency shift measurements were converted to force using the Sader-Jarvis method~\cite{Sader2004,Welker2012}. A stiffness of 1800~N/m for the frequency shift to force conversion was assumed. An additional electrode on the sensor was used to supply tunneling current. Tips were further sharpened by nitrogen etching while performing field ion microscopy~\cite{Rezeq2006a}. \textit{In-situ} tip processing was performed by controlled contacts of the tip to the sample surface which likely results in a decoration of the tip apex with silicon atoms~\cite{Huff2017,Labidi2017,Jarvis2012a}. Samples were cleaved from highly arsenic doped (1.5$\times$10$^{19}$ atom/cm$^{3}$) (100)-oriented Si crystals. After degassing at 600~$^\circ$C for ~12 hours, samples were flash annealed to temperatures as high as 1250~$^\circ$C before passivating the surface with hydrogen while maintaining a sample temperature of 330~$^\circ$C. The high flash temperatures have been previously shown to induce a dopant depletion region extending as far as 100~nm below the sample surface~\cite{Pitters2012,Rashidi2016}. DBs were patterned by applying short voltage pulses (+2.1~V, 10~ms) with the tip positioned directly above hydrogen~\cite{Lyding1994}. All tip offsets ($\Delta z$) used within the manuscript are in reference to an STM setpoint of -1.8~V and 50~pA measured over hydrogen.

\begin{figure}
	\includegraphics[width=85mm]{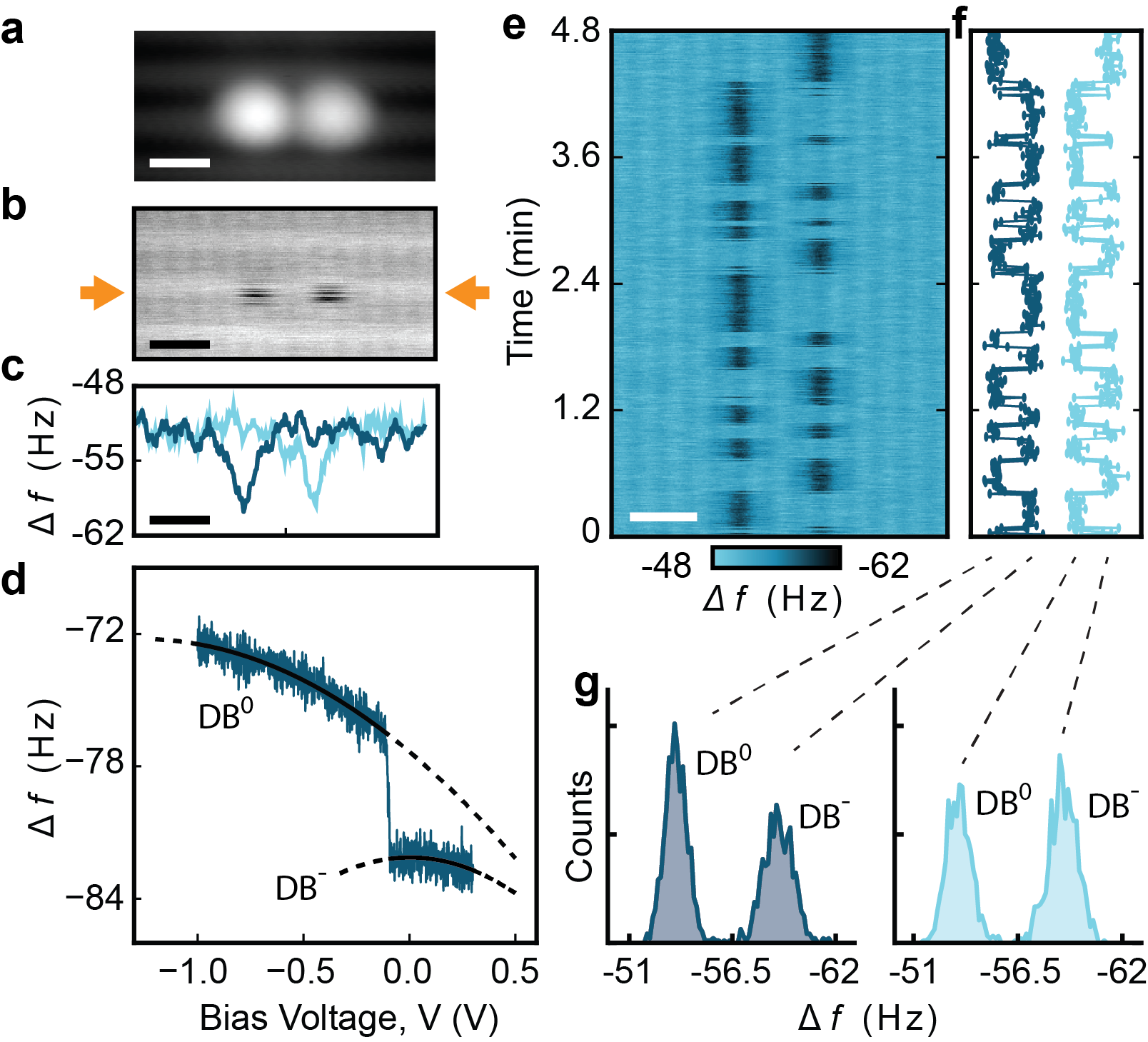}
	\caption{\label{Fig1} Charge configurations of two closely-spaced DBs. (a) Constant current filled state STM image, $-1.8$~V, 50~pA. (b) Constant height $\Delta f$ image, 0~V, $\-300$~pm. (c) Two constant height $\Delta f$ line scans (0~V, $-300$~pm) at the position indicated by the orange arrows in (b). (d) $\Delta f(V)$ spectroscopy taken above an isolated DB ($-370$~pm). The two individual segments have been fitted by two parabolas (solid lines: fit, dashed lines: extrapolation) corresponding to the neutral and negatively charged states (DB$^0$ and DB$^-$, respectively). (e) Combined map of 400 constant height $\Delta f$ line scans (0~V, $-300$~pm) taken sequentially over a 4.8 minute period. (f) Time-dependent bistable signal for the two individual DBs extracted from (e). (g) Histograms of the signals in (e). Labels indicate the charge state assignment of each peak. Scale bar is 1~nm (a-c,e).}
\end{figure}

In Figure~1, two DBs are patterned with two intervening hydrogen atoms using voltage pulses applied to the probe (Fig.~1a). Pairs of DBs are known to host only a single negative charge because the Coulombic repulsion between two closely-spaced negative charges would otherwise be too large~\cite{Haider2009};  here, constant height frequency shift ($\Delta f$) images of the pair appear streaky because the negative charge switches sites multiple times over the time it takes to acquire an image (Fig.~1b). This is seen clearly in individual $\Delta f$ line scans across the structure (Fig.~1c, taken at the position of the orange arrows in Fig.~1b) which reveal the localization of charge to one DB, with subsequent line scans demonstrating that this charge occasionally switches to the other DB. To definitively assign the change in contrast observed over each DB in $\Delta f$ images to a charge state, we performed bias-dependent $\Delta f$ spectroscopy ($\Delta f(V)$) on an isolated DB (Fig.~1d) which is negatively charged at 0~V on highly n-doped samples~\cite{Rashidi2017} (also see Supporting Information, Height-dependent contrast in $\Delta f$ images). It reveals a sharp transition between two parabolas~\cite{Gross2009}, associated with switching between the neutral (left of the step) and negatively charged states of the DB. By comparing the $\Delta f$ of the negatively charged state measured at 0~V to the fit of the neutral state's parabola at 0~V it becomes clear that the dark contrast (larger $\left|\Delta f\right|$) in Figure~1b,c corresponds to the negatively charged DB.

By stacking sequential $\Delta f$ line scans (Fig.~1f), we can monitor the charge switching between the two sites in real-time. Previous theoretical estimates for the tunneling rate between two closely-spaced DBs have ranged from THz to GHz, depending on the spacing~\cite{Livadaru2010,Shaterzadeh-yazdi2018}. Surprisingly, the bistable signal for each DB extracted from Fig.~1f demonstrates that the system's charge configuration often remains stable for seconds (Fig. 1g). Recent studies have revealed that charged species are often stabilized by a lattice relaxation of the supporting substrate~\cite{Repp2004,Olsson2007,Fatayer2018}. Density functional theory has similarly shown that negatively charged silicon DBs experience approximately 200~meV stabilization due to a relaxation of the lattice, which results in the nuclear position of the host atom being raised by approximately 30~pm relative to the neutral state~\cite{Northrup1989,Schofield2013,Kawai2016}. In this case, the lattice relaxation prevents the electron from elastically tunneling between the paired DBs. To assign the position of the charge in each $\Delta f$ line scan each trace was fitted with two Gaussian profiles. Histograms of the determined $\Delta f$ center values demonstrate two Gaussian profiles, representing the negative and neutral charge states of each DB (Fig.~1h and Supporting Information Fig.~S1). Because they are well separated, the charge state of each DB can be assigned reliably by a single line scan (Supporting Information Fig.~S3).

\begin{figure}
	\includegraphics[width=85mm]{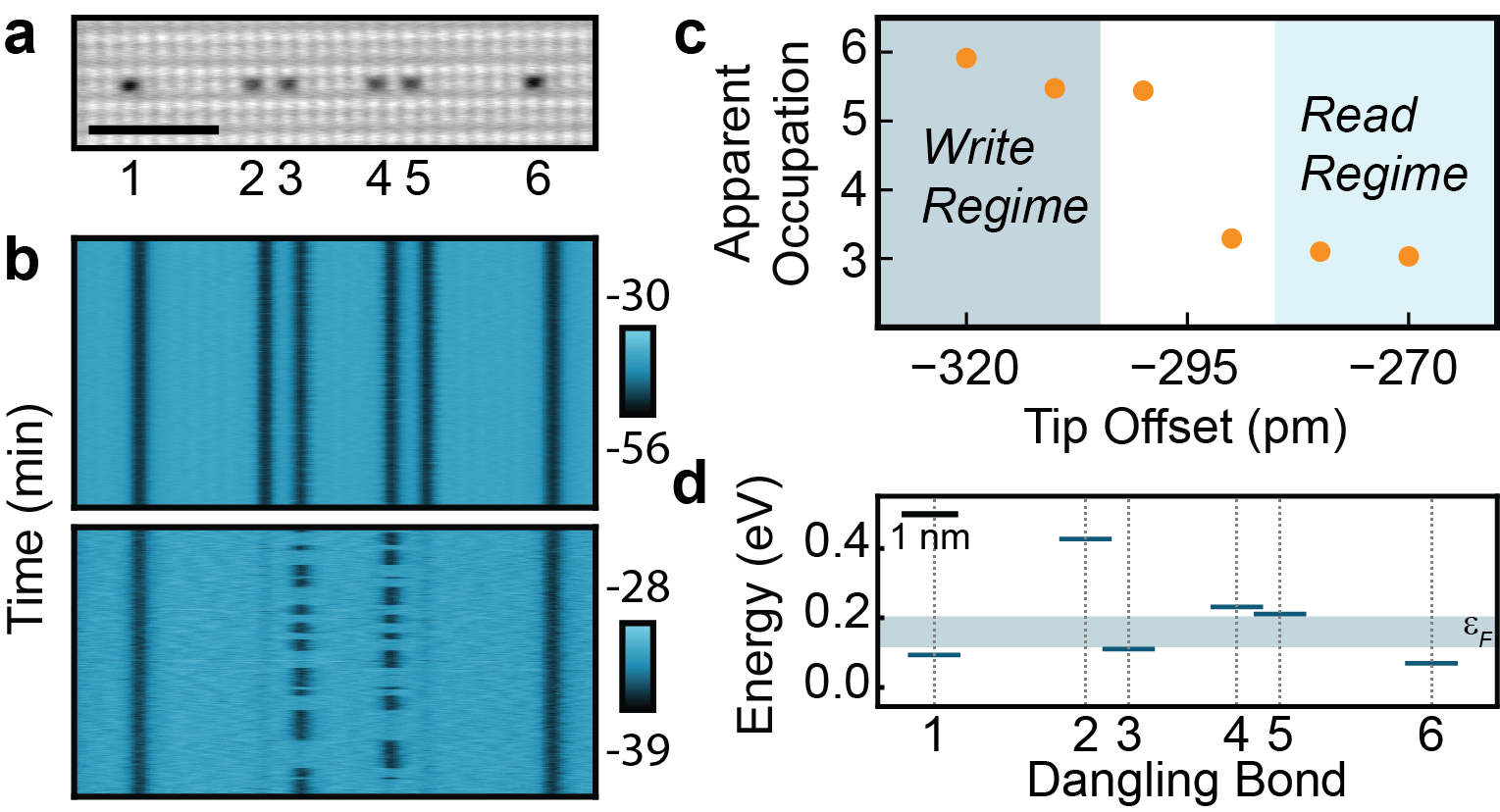}
	\caption{\label{Fig2}  Evolution of charge configurations of a symmetric six DB structure at different tip heights. (a) Constant height $\Delta f$ image (0 V, -300~pm). (b) Maps of 800 constant height $\Delta f$ line scans acquired over 18 minutes at -320~pm (top panel) and -270~pm (bottom panel). The scale bar in (a) is 3~nm and applies to (b). Color bars correspond to $\Delta f$. Histograms of the $\Delta f$ extracted over each DB in (b) are available in Supplementary Information Fig.~S3. (c) The average occupation of the structure inferred from digitizing the charge configuration at different $\Delta z$. Two interaction regimes: \textit{read} and \textit{write} are indicated. (d) Energetic shift of the neutral to negative charge transition (0/-) levels of each DB in the structure (1,3, and 6  are negatively charged) with respect to the (0/-) level of an isolated DB (0~eV). The levels are shifted Coulombically by the negative charges confined to the structure and are calculated in the absence of the probe using an electrostatic approximation of point charges and a surface dielectric constant of 6.35. Because the exact energy of the (0/-) level is unknown we indicate a range of energies (blue shaded area) over which the bulk Fermi level would give rise to the charge configurations observed in the lower panel of (b).
}
\end{figure}

Interestingly, we found that the occupation of DB structures observed at 0~V appeared to depend strongly on $\Delta z$. In Figure~2 we demonstrate this by performing a series of constant height line scan maps on a structure composed of six DBs with different $\Delta z$. The average occupation of each DB at each height can be inferred from the histograms of the $\Delta f$ measured over each DB (Supporting Information, Fig.~S3). More simply, the average occupation of the entire structure can be inferred by counting the number of dark bars in each line scan map. At the tip's closest approach (-320~pm, top panel Fig.~2b) all six DBs appear negatively charged. Upon withdrawing the tip by just 50~pm (-270~pm, Fig.~2b bottom panel) only three DBs are negatively charged at any given moment. This change in the apparent time-averaged occupation of the structure does not vary linearly with $\Delta z$, but instead transitions sharply between -300 and -290~pm (Fig.~2c).

To understand this trend we performed distance-dependent $\Delta f$ spectroscopy ($\Delta f(z)$) at 0~V on the individual DBs in a pair (Fig.~3a, blue curves) and over a vacancy on the surface (Fig.~3a, orange curve). We begin by withdrawing the tip 700~pm from our reference height to effectively eliminate the forces between the tip and sample and subsequently walk the tip towards the sample to progressively reintroduce them. Until approximately $\Delta z=-100$~pm all three curves are nearly identical, confirming that the long range forces (\textit{i.e.} van der Waals and capacitive forces due to the contact potential difference) are dominant~\cite{Lantz2001,Ternes2011,Sweetman2014}. Focusing on the approach curve obtained over the DB we note that at $\Delta z=-302\pm 2$ pm there is a sudden increase in the $\left|\Delta f\right|$  (observed at $\Delta z=-301\pm 2$ pm on the other DB). Crucially, we note that this results in hysteresis between the approach and retract curves, with the $\left|\Delta f\right|$ measured in the latter remaining larger until approximately $\Delta z=-100$~pm. Because of the similarity between the step in the approach curve and those observed in $\Delta f(V)$ experiments (e.g. Fig.~1d) we attribute this phenomenon to the localization of the pair's charge to the DB beneath the tip. Two observations confirm this: if a step is observed in the $\Delta f(z)$ obtained over one DB, subsequent $\Delta f(z)$ curves taken over the same DB do not demonstrate this behaviour. Instead, both the approach and retract curves trace the curve with the greater $\left|\Delta f\right|$, indicating the DB remains charged. On the other hand, if a step is observed in the $\Delta f(z)$ obtained over one DB and the subsequent $\Delta f(z)$ is performed on the other the hysteresis is consistently observed, indicating we  caused the charge to switch sites.

\begin{figure}
	\includegraphics[width=85mm]{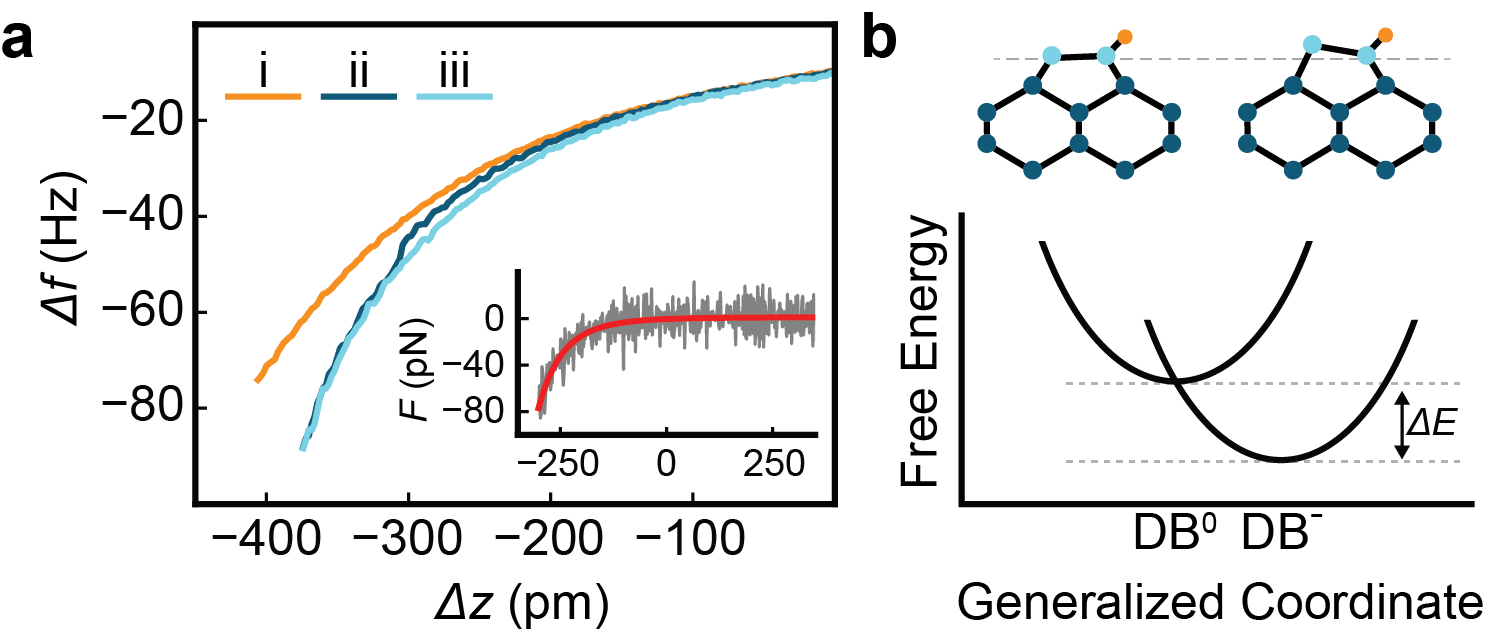}
	\caption{\label{Fig3}  Mechanically induced charge switching of a DB in a pair. (a) $\Delta f(z)$ at 0 V taken above an H atom (\textit{i}) and on a DB in a pair (\textit{ii} - approach; \textit{iii} - retract). Inset: the short range forces acting on the neutral DB (curve ii in (a)) up to the sharp step observed in $\left|\Delta f\right|$. See SI for more details. (b) Top panel: sketch of the equilibrium position of the host atom of a neutral DB (left) and a negatively charged DB (right). Dark blue, light blue, and orange atoms represent bulk silicon, silicon dimers,  and hydrogen, respectively.  Bottom panel: Free energy diagram depicting the neutral to negative charge transition for a DB due to the mechanical displacement of the host atom by the tip. $\Delta E$ corresponds to the lattice relaxation energy.}
\end{figure}

\begin{figure*}
	\includegraphics[width=130mm]{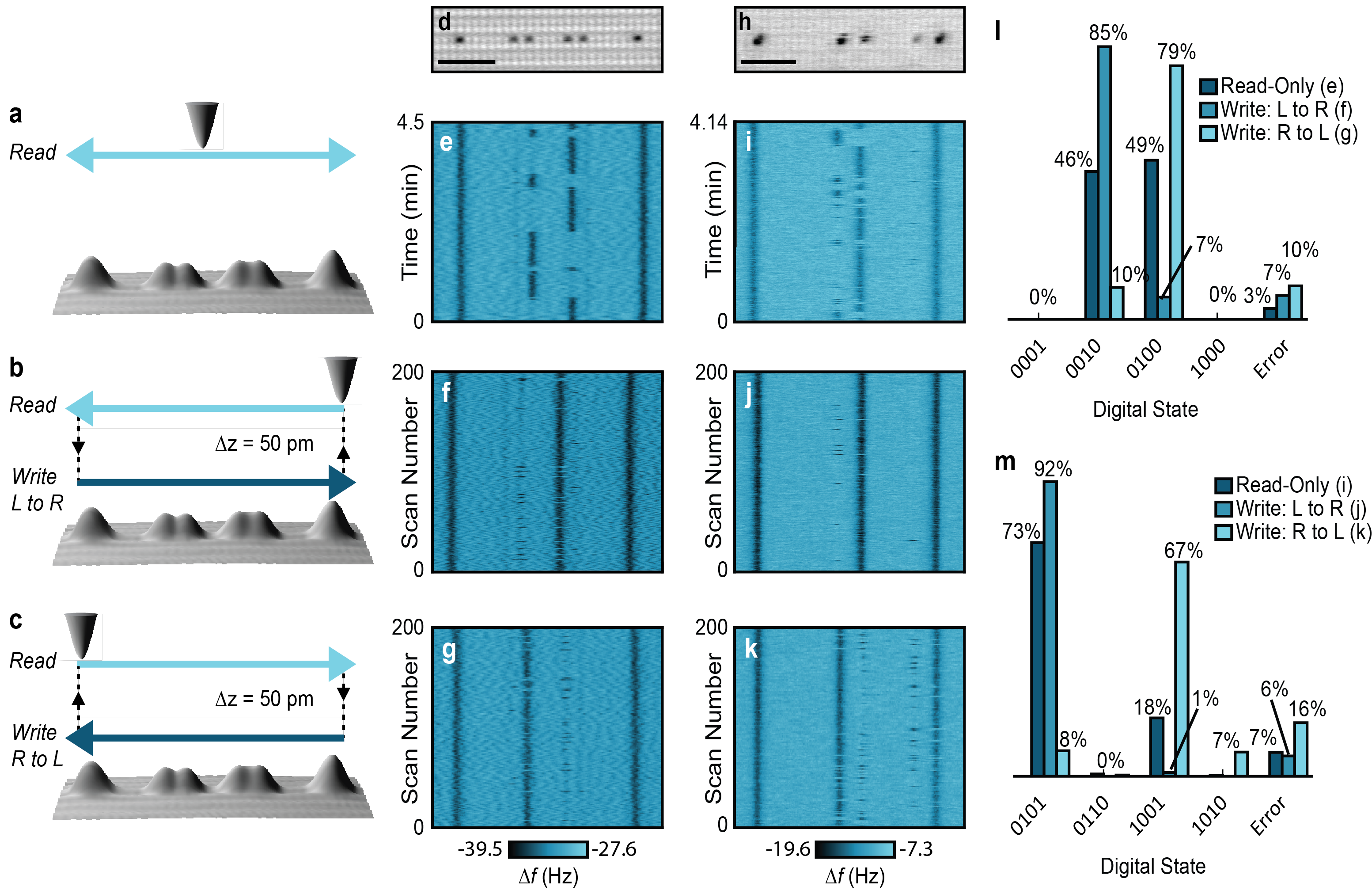}
	\caption{\label{Fig4}  Controlled preparation of charge configurations in symmetric and asymmetric DB structures. Visualization of the scan modes: (a) all measurements are restricted to the \textit{read regime}; (b) The tip is scanned from left to right (\textit{L to R}) in the \textit{write regime}, retracted 50~pm to the \textit{read regime}, and scanned back across; (c) the same process as (b) with directions reversed (\textit{write R to L}).  (d) Constant height $\Delta f$ image of the symmetric six-DB structure (0 V, -300 pm). (e-g) Maps of 200 line scans across structure in (d) corresponding to scheme (a) shown in (e), scheme (b) shown in (f), and scheme (c) shown in (g) (\textit{write regime}: -320~pm, \textit{read regime}: -270~pm). (h) Constant height $\Delta f$ image of the asymmetric five-DB structure, -350~pm, 0 V. (i-k) Maps of 200 line scans across structure (i) corresponding to scheme (a) (i), scheme (b) (j), and scheme (c) (k) (\textit{write regime}: -370~pm, \textit{read regime}: -320~pm). (l) Histograms of the binary numbers determined from digitization of the line scans in (e-g). 0's and 1's correspond to neutral and negatively charged DBs, respectively. Only the four interior DBs are considered. (m) Histograms of the binary numbers determined from digitization of the line scans in (i-k). The scale bars in (d) and (i) are 3~nm.
}
\end{figure*}
Similar hysteresis in $\Delta f(z)$ curves have been observed on the bare Si(100) surface previously~\cite{Sweetman2011}. In their case, the presence of  sudden hysteretic steps corresponded directly to a toggling of the buckling direction of a single Si(100) dimer on the hydrogen-free surface. The authors concluded that at small absolute tip heights short range forces between the probe and sample result in a mechanically-induced deformation of the lattice. The same mechanism is at play in our experiments. One distinction of our work is that the mechanical deformation also corresponds to a change in the charge state of the surface atom. This can be understood by considering the equilibrium positions of the host silicon atom for a negative and a neutral DB, which as noted earlier, differ due to relaxation of the lattice (sketch in top panel, Fig.~3b)~\cite{Northrup1989,Schofield2013,Kawai2016}. Because the forces are all attractive at the height corresponding to the step in the approach curve, a displacement of the surface atom is expected to be in the direction of the tip, causing the atom to re-hybridize and adopt greater $sp^3$ character. Consequently, the total free energy of the negatively charged state is lowered with respect to the neutral state, leading to the charging of the DB beneath the tip (bottom panel, Fig.~3b). Using $\Delta f(z)$ curves obtained over dimer vacancies on the surface (Supporting Information Fig. S4) we separate the long and short range force contributions~\cite{Lantz2001,Ternes2011,Sweetman2014}. We find the short range forces required to lift the equilibrium position of the neutral host atom is fit best by a function of the form $-\mathrm{C}/z^{7}$, where $z$ is the absolute tip height and C is a constant (Supporting Information Fig. S6). This strongly suggests that van der Waals forces are responsible for displacing the host atom~\cite{Bocquet2011}. A force of $-75 \pm 13$~pN ($-77 \pm 12$~pN) was found for the right (left) DB (inset Fig.~3a, uncertainty in force measurement corresponds to one standard deviation) which is consistent with the force reported by Sweetman et. al. required to toggle the Si dimer~\cite{Sweetman2011,Jarvis2012a}.

The experiments in Figure~2 can now be clearly explained. At small absolute tip heights the short range forces are strong enough that as the probe scans over the structure the charging of each DB becomes favorable whenever it is beneath the tip (top panel, Fig.~2b). This necessitates that electrons vacate prior negatively charged DBs such that the overall occupation of the structure remains constant. Upon withdrawing the tip a short distance, however, this effect is greatly diminished. As a result we observe that specific charge configurations can remain stable for many sequential measurements ($>15$ s on average, bottom panel, Fig.~2b). It is also interesting to note that only two charge configurations appear consistently: the two outer DBs remain continuously charged and a single negative charge is observed to switch between the two central DBs, similar to the behavior observed on an isolated pair. By observing that the total amount of time the central charge spends in the left DB (50$\%$) is roughly equal to the right (46$\%$), and noting the structure's symmetry, it is clear that these two charge configurations correspond to the degenerate ground state. We do not observe higher energy charge configurations of this structure, likely because the Coulombic interaction between closely spaced negative charges makes them energetically unfavorable, \textit{e.g.}, if DBs 1, 2, and 6 in Fig. 2d were negatively charged. We therefore identify two interaction regimes (Fig.~2c): one where charge can be controllably manipulated by the tip (the \textit{write regime}) and another where we can observe stable or metastable charge configurations (the \textit{read regime}).

To further validate our assignment of the \textit{write} and \textit{read} regimes we performed the experiments depicted in schemes Figures~4a-c on the symmetric structure (Fig.~4d) and an asymmetric structure composed of five DBs (Fig.~4h). First, we restricted the measurements to the \textit{read regime} (scheme Fig.~4a, Fig.~4e,i), allowing us to characterize the intrinsic charge configurations of the structures and assess their relative energies based on how often they occur (dark blue in Fig.~4l,m). Subsequent experiments contain two associated phases: in the \textit{write phase}, the tip is scanned across the structure at close proximity; in the \textit{read phase}, the tip is retracted 50~pm with respect to the \textit{write phase} and scanned back across (depicted in schemes Fig.~4b,c). It might be expected that any charge configuration prepared by the \textit{write phase} should be observed in the \textit{read phase}. Indeed, Figure~4f,g and 4j,k confirm that charge in the interior of both structures can be manipulated. On the symmetric structure we could consistently initiate charge to the right (Fig.~4f, 85$\%$) or left (Fig.~4g, 79$\%$) central DB, corresponding to preparation of the degenerate ground state configurations observed in Figure~4e. On the asymmetric structure measurements restricted to the \textit{read regime} (Fig.~4i) demonstrate that this system has three negative charges. On this structure also only the charge confined to the inner pair fluctuates, but because the structure is asymmetric these two charge configurations are non-degenerate. Although we expected the interior charge to favor the left DB of the pair we observe the opposite (Fig.~4i,m 18$\%$ \textit{vs.} 73$\%$, respectively). This indicates that other charged species (\textit{e.g.} DBs or ionized donors) act as an additional electrostatic bias on this structure. We note, however, that these hidden biases can be counteracted by patterning additional DBs in the structure's surrounding area (Supporting Information, Fig.~S5). Using the techniques previously described the central charge could be manipulated to selectively occupy the right (Fig.~4j, 92$\%$) or left (Fig.~4k. 67$\%$) DB of the pair, demonstrating that in addition to the ground state configurations the occurrence of metastable charge configurations can also be enhanced (Fig.~4i).

These results demonstrate that single electrons can be sensed and manipulated within structures derived from DBs. We find that charge configurations can remain stable for periods on the order of several seconds as a result of a relaxation of the silicon lattice which acts to stabilize negatively charged DBs. Using the probe, specific charge configurations can be prepared. The mechanism used to achieve this does not depend on the application of bias voltage--instead it is due to a mechanical manipulation of the DB's host atom with the probe which initiates a change of its charge state. The techniques presented here expand the scanning probe toolkit with the ability to position charge within atomic structures and prepare desired charge configurations.

\begin{acknowledgments}
We would like to thank Mark Salomons and Martin Cloutier for their technical expertise. We also thank Leo Gross and Gerhard Meyer for stimulating discussions. We thank NRC, NSERC, and AITF for their financial support.

\end{acknowledgments}

M.R., W.V. and T.D. contributed equally to this work.

\clearpage
\onecolumngrid
\appendix
\includegraphics[width=18cm,page=1]{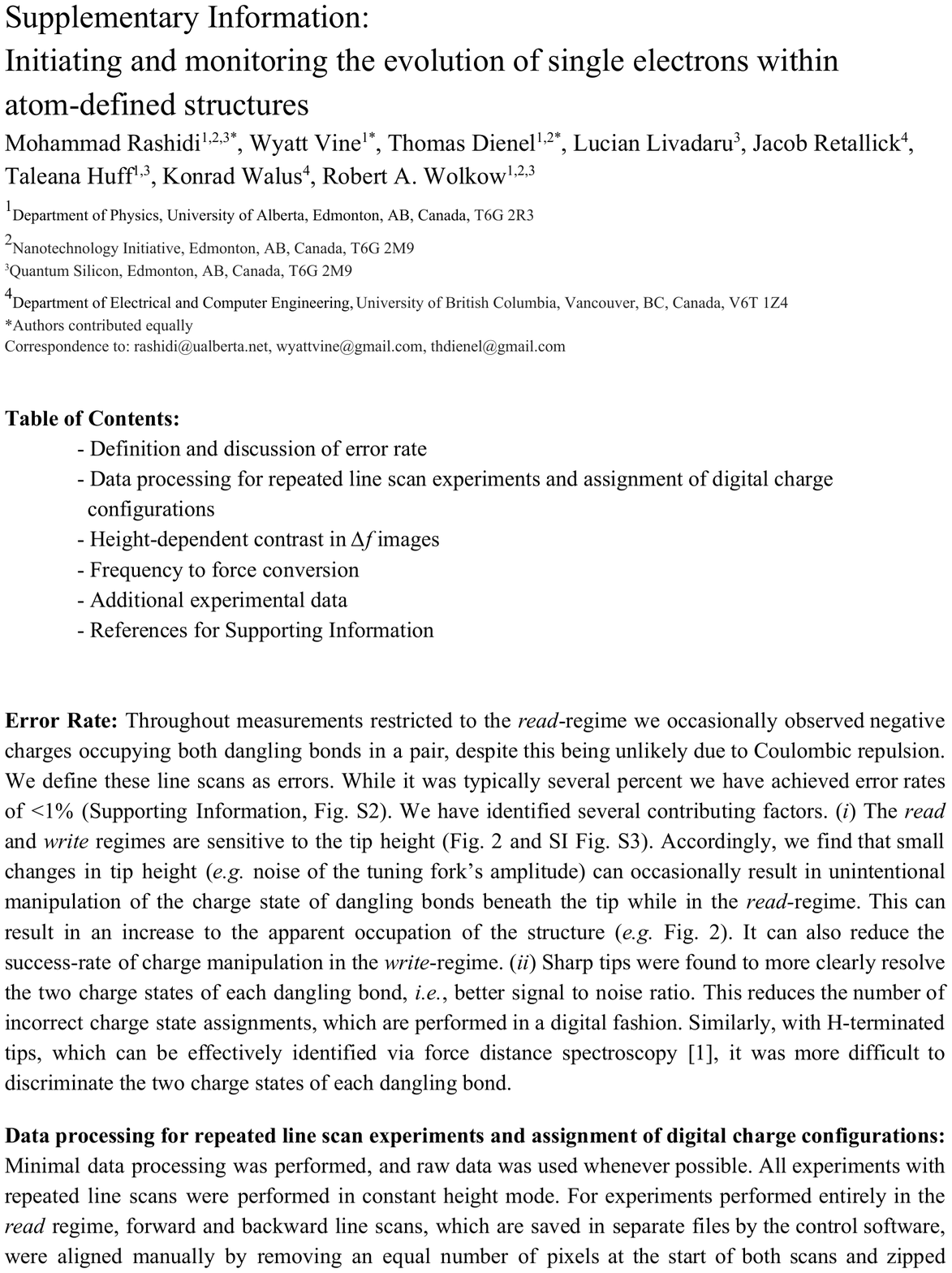}
\includegraphics[width=18cm,page=2]{Supplementary_Information.pdf}
\includegraphics[width=18cm,page=3]{Supplementary_Information.pdf}
\includegraphics[width=18cm,page=4]{Supplementary_Information.pdf}
\includegraphics[width=18cm,page=5]{Supplementary_Information.pdf}
\includegraphics[width=18cm,page=6]{Supplementary_Information.pdf}
\includegraphics[width=18cm,page=7]{Supplementary_Information.pdf}
\end{document}